\documentclass[12pt]{article}
\usepackage {graphicx}
\usepackage{textcomp}

\begin{document}

\rightline{  }

\vspace*{1.5cm}

\begin{center}

\LARGE{The MSSM with Large Gluino Mass \\[20mm]}

\large{Yu.M.Andreev, N.V.Krasnikov  and  A.N.Toropin}

\it{Institute for Nuclear Research RAS\\
Moscow, 117312, Russia\\[15mm]}

\large{\rm{Abstract}} \\[7mm]

\end{center}

\begin{center}

\begin{minipage}[h]{14cm}

We study the Minimal Supersymmetric Standard Model (MSSM) with large gluino 
mass $m_{\tilde{g}} \gg 1~TeV$.
In particular, we discuss the LHC supersymmetry discovery signatures with 
$n ~leptons ~+~ jets ~+~E^{miss}_T,  ~n\geq0 $ for the MSSM with large gluino 
mass. We show that for some relations among squark and neutralino masses 
leptonic signatures with $n ~leptons ~+~ jets ~+~E^{miss}_T,~n\geq1$ do not 
allow to discover supersymmetry at the LHC and the only supersymmetry discovery 
signature remains the signature with $no ~leptons ~+~ jets ~+~E^{miss}_T$. 
Moreover, for LSP mass close to squark masses the LHC discovery potential for 
this signature is strongly reduced. 

\end{minipage}

\end{center}

\newpage

\pagestyle{plain}

One of the supergoals of the Large Hadron Collider (LHC) \cite{1} is the 
discovery of supersymmetry \cite{2}. The simplest supersymmetric 
generalization of the Standard Model (SM) \cite{SM} is the Minimal 
Supersymmetric Standard Model (MSSM) \cite{SUSY}. In the MSSM all sparticle 
masses are arbitrary that complicates the analysis. In the Minimal 
Supergravity model (mSUGRA) \cite{SUSY} the universality of the different 
soft mass parameters at the Grand Unified Theory (GUT) scale 
$M_{GUT} \approx 2 \cdot 10^{16}~GeV$ is postulated. The renormalization 
group equations are used to relate GUT and electroweak scales. The equations 
for the determination of nontrivial minimum of the electroweak potential are 
used to decrease the number of unknown parameters by two. So mSUGRA model 
depends on five unknown parameters. Despite the simplicity of the mSUGRA 
model it is a very particular model. Moreover at present there are some 
string inspired models with nonuniversal sfermion and gaugino masses \cite{NIL}
\footnote{
Phenomenological consequences of the MSSM with non mSUGRA spectrum were studied 
in Refs \cite{KR1}, \cite{nonKR2}, \cite{nonKR3}.
}. 
One loop quadratic correction to the Higgs boson mass at the SM is given by the 
formula \cite{Barb}  
\begin{equation}
\delta m^2_h = \alpha_t \Lambda_t^2 + \alpha_g \Lambda_g^2 +
\alpha_h \Lambda_h^2  \,,
\end{equation}
where 
\begin{equation}
\alpha_t = \frac{3m^2_t}{4\pi^2v^2}, ~\alpha_g = -\frac{6m^2_W +3m^2_Z}
{16\pi^2v^2}, ~\alpha_h = -\frac{3m^2_h}{16\pi^2v^2}
\end{equation}
and $\Lambda_i$ are the ultraviolet cutoffs of the momenta of virtual top 
quarks, gauge bosons, and the Higgs boson itself. From the naturalness 
condition $\delta m^2_h \leq  m^2_h$ one can find that 
$\Lambda_i \leq O(1)~TeV$. One loop naturalness condition (1) predicts that 
the masses of stop quarks and supersymmetric analogs of electroweak gauge 
bosons $W$ and $Z$ have to be lighter than $O(1)~TeV$. It should be stressed 
that gluon corrections to the Higgs boson mass arise only at two loop level 
and as a consequence from the naturalness condition gluino mass has to be 
lighter than $O(10)~TeV$, i.e. it could be much higher than $1~TeV$. 
Also from the naturalness point of view supersymmetric analogs of the first 
and the second generation quarks and leptons can have masses $O(10)~TeV$ 
since the corresponding Yukawa couplings are small.

In this note we study the MSSM with large gluino mass 
$m_{\tilde{g}} \gg m_{\tilde{q}}$
\footnote{
Note that the MSSM with large gluino mass is an opposite case to the `focus 
point` mSUGRA scenario \cite{Focus} with $m_0 \gg m_{1/2}$ (squark and slepton 
masses are large).
}.
Namely, we study the LHC supersymmetry discovery signatures with 
$n \geq 0 ~leptons ~+~ jets ~+~E^{miss}_T$ for the MSSM model with large 
gluino mass. We show that for some relations among squark and gluino masses, 
the LHC signatures with $n \geq 1 ~leptons ~+~ jets ~+~E^{miss}_T$ do not 
allow to discover supersymmetry and the only supersymmetry discovery signature 
remains the signature with $no ~leptons ~+~ jets ~+~E^{miss}_T$. Moreover, for 
LSP mass close to squark masses, LHC supersymmetry discovery potential for 
this signature is strongly reduced.

The gluino and squark production cross sections are the biggest ones compared 
to slepton or gaugino cross sections. Therefore gluinos and squarks production 
at the LHC are the most interesting reactions from the supersymmetry discovery 
point of view with the cross sections around $1~pb$ for squark and gluino 
masses around $1~TeV$. The squark and gluino decays produce missing transverse 
energy from the LSP plus multiple jets and varying numbers of leptons from the 
intermediate gauginos. It is natural to divide the signatures used for the 
squark and gluino detections into the following two groups:

a. $no ~leptons ~+~ jets ~+~E^{miss}_T$ events,

b. $n  ~leptons ~+~ jets ~+~E^{miss}_T,~n\geq1$ events.

Multileptons arise as a result of the cascade decays of neutralinos and 
charginos into $W$- and $Z$-bosons with subsequent decays of $W$- and 
$Z$-bosons into leptonic modes. For instance, the same sign and opposite sign 
dilepton events arise as a result of the cascade decay
\begin{equation}
\tilde{g} \rightarrow q^{'} \bar{q} \tilde{\chi}^{\pm}_{i}, \; 
\tilde{\chi}^{\pm}_{i}
\rightarrow W^{\pm}\tilde{\chi}^{0}_1  \rightarrow l^{\pm} 
\nu \tilde{\chi}^0_1 \,,
\end{equation}
where $l$ stands for both $e$ and $\mu$. Opposite sign dilepton events can 
arise also as a result of cascade decay
\begin{equation}
\tilde{g} \rightarrow q \bar{q} \tilde{\chi}^{0}_{i} , \;
\tilde{\chi}^0_i \rightarrow Z \tilde{\chi}^0_1 \rightarrow
l^{+}l^{-} \tilde{\chi}^0_1 \,.
\end{equation}
The main conclusion \cite{TDR} is that for the mSUGRA model the LHC (CMS) 
will be able to discover supersymmetry with squark or gluino masses up to 
$(2 - 2.5)~TeV$ for $L_{tot} = 30~fb^{-1}$. The most powerful signature for 
squark and gluino detection in the mSUGRA model is the signature with 
multijets and $E_T^{miss}$ (signature a).

Note that the branchings of the decays $\tilde{\chi}^0_2 \rightarrow l^+l^- 
\tilde{\chi}^0_1$, $\tilde{\chi_1}^{\pm} \rightarrow l^{\pm} \nu 
\tilde{\chi}^0_1$ leading to the signatures with leptons depend on the 
relation among $m_{\tilde{\chi}^0_2}$ and $m_{\tilde{\chi}^0_1}$. For some 
relations among supersymmetry masses the branchings into leptons are 
suppressed and as a consequence the signatures with leptons do not allow to 
discover supersymmetry. 

In this note we considered the modified CMS test point LM1
\footnote{
For mSUGRA test point LM1 $m_0 = 60~GeV, m_{1/2} = 250~GeV, \tan\beta = 10, 
A = 0, sign \mu  = +$.
}.
Namely, we considered the following modifications in mass spectrum of the 
point LM1
\footnote{
For test point LM1 the masses of some supersymmetric particles are:
$m_{\tilde{\chi}^0_1} = 99.6~GeV$,
$m_{\tilde{\chi}^0_2} = 186.4~GeV$, 
$m_{\tilde{g}} = 579.5~GeV$, 
$m_{\tilde{e}_L} = 195.7~GeV$, 
$m_{\tilde{e}_R} = 122~GeV$, 
$m_{\tilde{U}_L} = 554~GeV$, 
$m_{\tilde{D_R}} = 530~GeV $.
}:
 
a1. $ m_{\tilde g} \rightarrow 3 m_{\tilde g}$;

b1. $ m_{\tilde g} \rightarrow 3 m_{\tilde g}, 
   ~m_{\tilde{\chi}^0_1} \rightarrow  3 m_{\tilde{\chi}^0_1}  $;

c1. $ m_{\tilde g} \rightarrow 3 m_{\tilde g}, 
   ~m_{\tilde{\chi}^0_2} \rightarrow  3 m_{\tilde{\chi}^0_2}  $;

d1. $ m_{\tilde g} \rightarrow 3 m_{\tilde g}, 
   ~m_{\tilde{\chi}^0_2} \rightarrow  3 m_{\tilde{\chi}^0_2}, 
   ~\mu \rightarrow 3 \mu $;

e1. $ m_{\tilde g} \rightarrow 3 m_{\tilde g}, 
   ~m_{\tilde{\chi}^0_1} \rightarrow  3 m_{\tilde{\chi}^0_1}, 
   ~m_{\tilde{\chi}^0_2} \rightarrow  3 m_{\tilde{\chi}^0_2}, 
   ~\mu \rightarrow 3 \mu $.

f1. $ m_{\tilde g} \rightarrow 3 m_{\tilde g}, 
   ~m_{\tilde{q}_{1,2}} \rightarrow  3 m_{\tilde{q}_{1,2}}, 
   ~m_{\tilde{l}_{1,2}} \rightarrow  3 m_{\tilde{l}_{1,2}}$. 
 
We also considered test points with squark and slepton masses $m_{\tilde{q}} = 
m_{\tilde{l}} = 800 ~GeV$, gluino mass $m_{\tilde{g}} = 3000~GeV$ and gaugino masses:

a2. $m_{\tilde{\chi}^0_1} = 100~GeV, ~m_{\tilde{\chi}^0_2} = 400~GeV, ~\mu = 400~GeV$;

b2. $m_{\tilde{\chi}^0_1} = 100~GeV, ~m_{\tilde{\chi}^0_2} = 900~GeV, ~\mu = 400~GeV$;

c2. $m_{\tilde{\chi}^0_1} = 100~GeV, ~m_{\tilde{\chi}^0_2} = 900~GeV, ~\mu = 800~GeV$;

d2. $m_{\tilde{\chi}^0_1} = 300~GeV, ~m_{\tilde{\chi}^0_2} = 900~GeV, ~\mu = 800~GeV$;

e2. $m_{\tilde{\chi}^0_1} = 600~GeV, ~m_{\tilde{\chi}^0_2} = 900~GeV, ~\mu = 800~GeV$;

f2. $m_{\tilde{\chi}^0_1} = 700~GeV, ~m_{\tilde{\chi}^0_2} = ~900~GeV, ~\mu = 800~GeV$.

The coupling constants and cross sections for SUSY processes were calculated 
with PYTHIA code \cite{PYT}. We used the full simulation results of Ref.\cite 
{TDR} for the estimation of  background events. The CMS simulation codes CMSSW 
\cite{CMSSW} in fast mode were used.  

For the signature with two opposite charge and the same flavour leptons:
$l^+l^- ~+~E^{miss}_T$ we have used the following selection cuts \cite{TDR}:

cut on leptons: $p^{lept}_T  > 20~GeV$, $|\eta| < 2.4$, 
lepton isolation within $\Delta R < 0.3$ cone,

cut on missing transverse energy: $E^{miss}_T > 300~GeV$. 

For integrated luminosity $L_{tot} = 10~fb^{-1}$ the number of background events 
is $N_{bkg} = 93$ \cite{TDR}. For points LM1 and (a1 - f1) the number of signal 
events and the significance $S_{c12} = 2(\sqrt{N_{sig}+N_{bkg}}-\sqrt{N_{bkg}})$ 
\cite{KB} are presented in Table 1.

\begin{table}[htb]
\begin{center}
\begin{tabular}{ccccc}
\hline
\hline
Point &  $N_{sig}$ & $S_{c12}$ \\
\hline
LM1   &   91  & 13.9   \\
a1    &   15  &  2.1   \\
b1    &    7  &  1.0   \\
c1    &    3  &  0.4   \\
d1    &   29  &  3.8   \\
e1    &    4  &  0.6   \\
f1    &    1  &  0.1   \\
\hline
\hline
\end{tabular}
\caption{
The number of signal events and significance $S_{c12}$ for points LM1 and 
(a1 - f1) at $L_{tot} = 1~fb^{-1}$, dilepton signature $l^+l^- ~+~E^{miss}_T $.
}
\end{center}
\end{table}

\begin{table}[htb]
\begin{center}
\begin{tabular}{ccccc}
\hline
\hline
Point &  $N_{sig}$ & $S_{c12}$ \\
\hline
LM1   &  6319  &  130.8  \\
a1    &  1001  &   23.8  \\
b1    &   853  &   20.6  \\
c1    &   693  &   18.1  \\
d1    &   926  &   22.0  \\
e1    &   669  &   16.7  \\
f1    &    83  &    3.4  \\
\hline
\hline
\end{tabular}
\caption{
The number of signal events and significance $S_{c12}$ for points LM1 and 
(a1 - f1) at $L_{tot} = 1~fb^{-1}$, signature $no~leptons ~+~ n \geq ~3~jets 
~+~E^{miss}_T $.
}
\end{center}
\end{table}

\begin{table}[htb]
\begin{center}
\begin{tabular}{ccccc}
\hline
\hline
Point &  $N_{sig}$ & $S_{c12}$ \\
\hline
a2    &  9.6  &   1.4   \\
b2    &  4.9  &   0.7   \\
c2    &  3.9  &   0.6   \\
d2    &  0.2  &   0.03  \\
e2    &   0   &   0     \\
f2    &   0   &   0     \\
\hline
\hline
\end{tabular}
\caption{
The number of signal events and significance $S_{c12}$ for points (a2 - f2) at 
$L_{tot} = 1~fb^{-1}$, dilepton signature  $ l^+l^- ~+~E^{miss}_T $.
}
\end{center}
\end{table}

\begin{table}[htb]
\begin{center}
\begin{tabular}{ccccc}
\hline
\hline
Point &  $N_{sig}$ & $S_{c12}$ \\
\hline
a2    &  185   &  10.2   \\
b2    &  145   &   8.2   \\
c2    &  141   &   8.0   \\
d2    &   68   &   4.1   \\
e2    &   23.5 &   1.5   \\
f2    &    8.5 &   0.5   \\
\hline
\hline
\end{tabular}
\caption{
The number of signal events and significance $S_{c12}$ for points (a2 - f2) at 
$L_{tot} = 1~fb^{-1}$, signature $no ~leptons ~+~ n \geq 3 ~jets  ~+~E^{miss}_T $.
}
\end{center}
\end{table}

As we see from Table 1 only for points LM1 and a1 we can expect supersymmetry 
discovery at $L_{tot} = 1~fb^{-1}$. Moreover even for the larger luminosities 
our conclusion does not change. As we checked the use of the signature  
$1~lepton + ~jets~ + E_T^{miss}$ does not help to discover supersymmetry and 
the only signature for supersymmetry discovery remains the signature with 
$no~leptons +~jets~ + E_T^{miss}$. For this signature we used cuts \cite{TDR}:

$ E^{miss}_T \geq 200~GeV$, $n_{jet} > 3$, 

$E_{T,j1} \geq 180~GeV$, $ E_{T,j2} \geq 110~GeV$, $E_{T,j3} \geq 30~GeV$,

$E^{miss}_T + E_{T,j2}+ E_{T,j3} + E_{T,j4} \geq 500~GeV $.

We have found that for the signature  $no~leptons + ~jets~ + E_T^{miss}$ the 
use of these cuts allows to discover supersymmetry for points (a1 - e1)  at 
the integrated luminosity $L_{tot} = 1~fb^{-1}$. 

For points (a2 - f2) we checked that the signature with $n \geq 1 ~leptons 
+ ~jets~ + E_T^{miss}$ does not allow to discover supersymmetry even at high 
integrated luminosity $L_{tot} = 100~fb^{-1}$. For instance, for point b2 the 
number of signal events at $L_{tot} = 1~fb^{-1}$ is $N_{sig} = 2$ (see Table 3). 
The reason is that the supersymmetry cross section $\sigma(pp \rightarrow 
sparticles + ...) \approx 0.7~pb$ is not very big and cascade decays $\tilde{q} 
\rightarrow \tilde{\chi}^0_2 q \rightarrow l^+l^- \tilde{\chi}^0_1$, $\tilde{q} 
\rightarrow \tilde{\chi}^{\pm}_2 q\prime \rightarrow l^{\pm}\nu \tilde{\chi}^0_1$,
leading to the signatures with multileptons in final state are suppressed. The 
main decay mode becomes decay $\tilde{q} \rightarrow q \tilde{\chi}^0_1$ that 
means in particular that in the signature $no ~leptons + ~jets~ + E^T_{miss}$
the events with two hadron jets dominate. For the signature $no ~leptons + 
~jets~ + E_T^{miss}$ we found that it is possible to discover supersymmetry for 
points (a2 - f2). For point f2 we found that the use of the cuts 
\cite{TDR} does not allow to discover supersymmetry. The reason is that for 
$m_{\tilde{\chi}^0_1}$ close to $m_{\tilde{ q}}$ the decay $\tilde{q} 
\rightarrow q \tilde{\chi}^0_1$ leads to more soft distributions in $E^{miss}_T$ 
and $E_{T,jet}$ compared to the case when $m_{\tilde{q}} \gg m_{\tilde{\chi^0_1}}$ 
\cite{KR1} so the number of signal events is decreased.

To conclude, in this note we studied the MSSM with large gluino mass $m_{\tilde{g}} 
\gg m_{\tilde{q}}$. Namely, we investigated the LHC supersymmetry discovery 
signatures with $n \geq 0 ~leptons ~+~ jets ~+~E^{miss}_T$ for the MSSM with 
large gluino mass. We found that for some relations among squark and 
gaugino masses signatures with $n \geq 1 ~leptons ~+~ jets ~+~E^{miss}_T$ 
isolated leptons do not allow to discover supersymmetry and the only 
supersymmetry discovery signature remains the signature $no ~leptons ~+~ jets 
~+~E^{miss}_T$. Moreover, for LSP mass close to squark masses the LHC 
supersymmetry discovery potential for this signature is strongly reduced. 

We are indebted to V.A.Matveev for valuable comments. 
This work has been supported by the RFFI grants \textnumero 07-02-00256 and 
\textnumero 08-02-91007.

\newpage

\begin{figure}[!htbp]
  \begin{center}
    \resizebox{11.5cm}{!}{\includegraphics{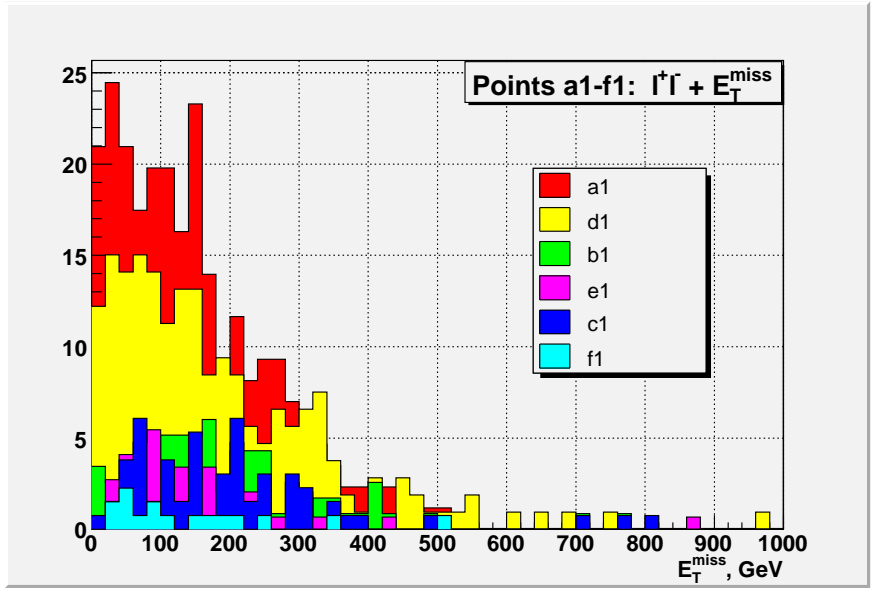}}
    \caption{
             The distributions of $E^{miss}_T$ for points a1 - f1,
             dilepton signature  $ l^+l^- ~+~E^{miss}_T $.
            }
    \label{a1f1dilepton} 
  \end{center}
\end{figure}

\begin{figure}[!htbp]
  \begin{center}
    \resizebox{11.5cm}{!}{\includegraphics{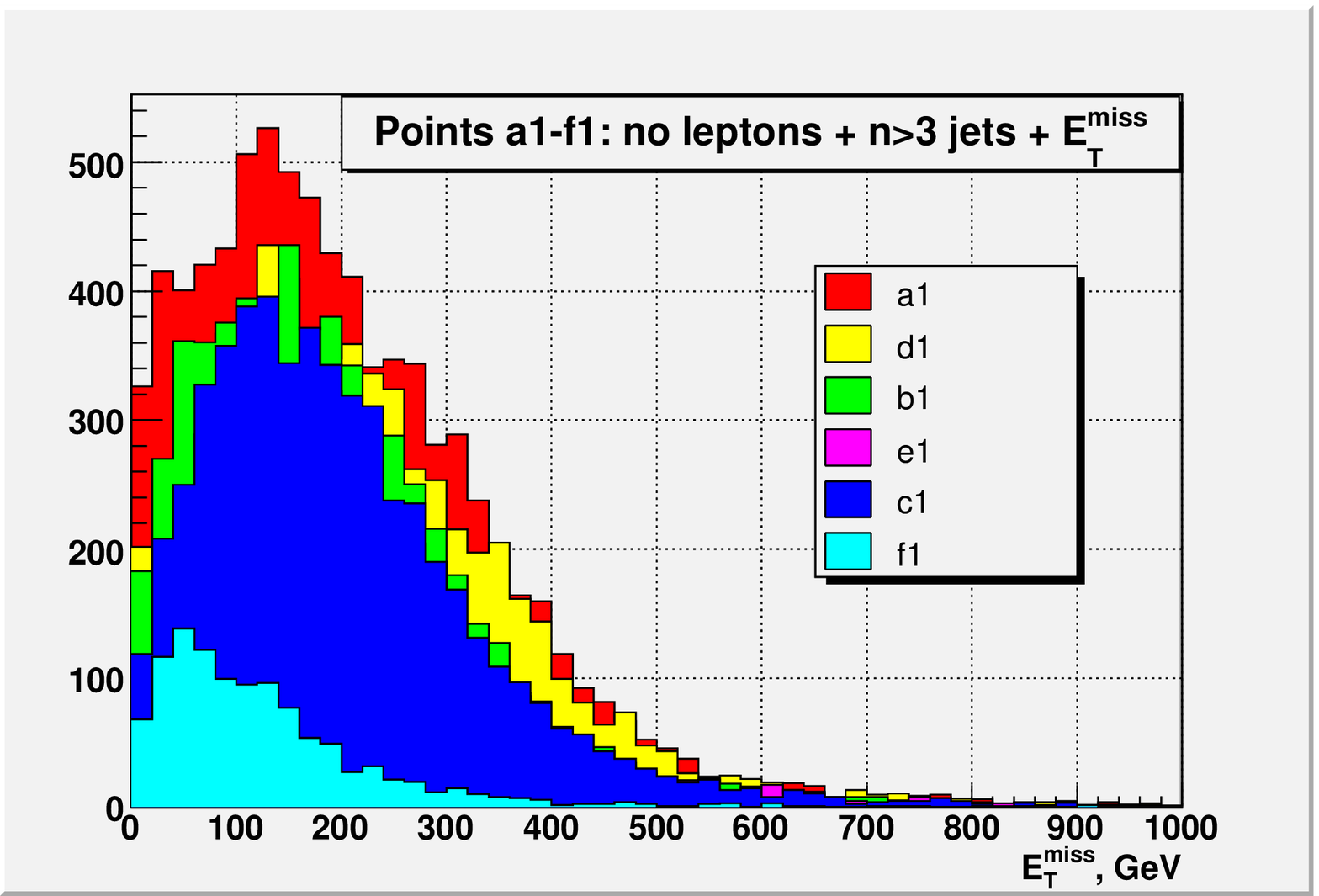}}
    \caption{
             The distributions of $E^{miss}_T$ for points a1 - f1,
             signature $no~leptons ~+~ n \geq ~3~jets ~+~E^{miss}_T $.
            }
    \label{a1f1jets} 
  \end{center}
\end{figure}

\begin{figure}[!htbp]
  \begin{center}
    \resizebox{11.5cm}{!}{\includegraphics{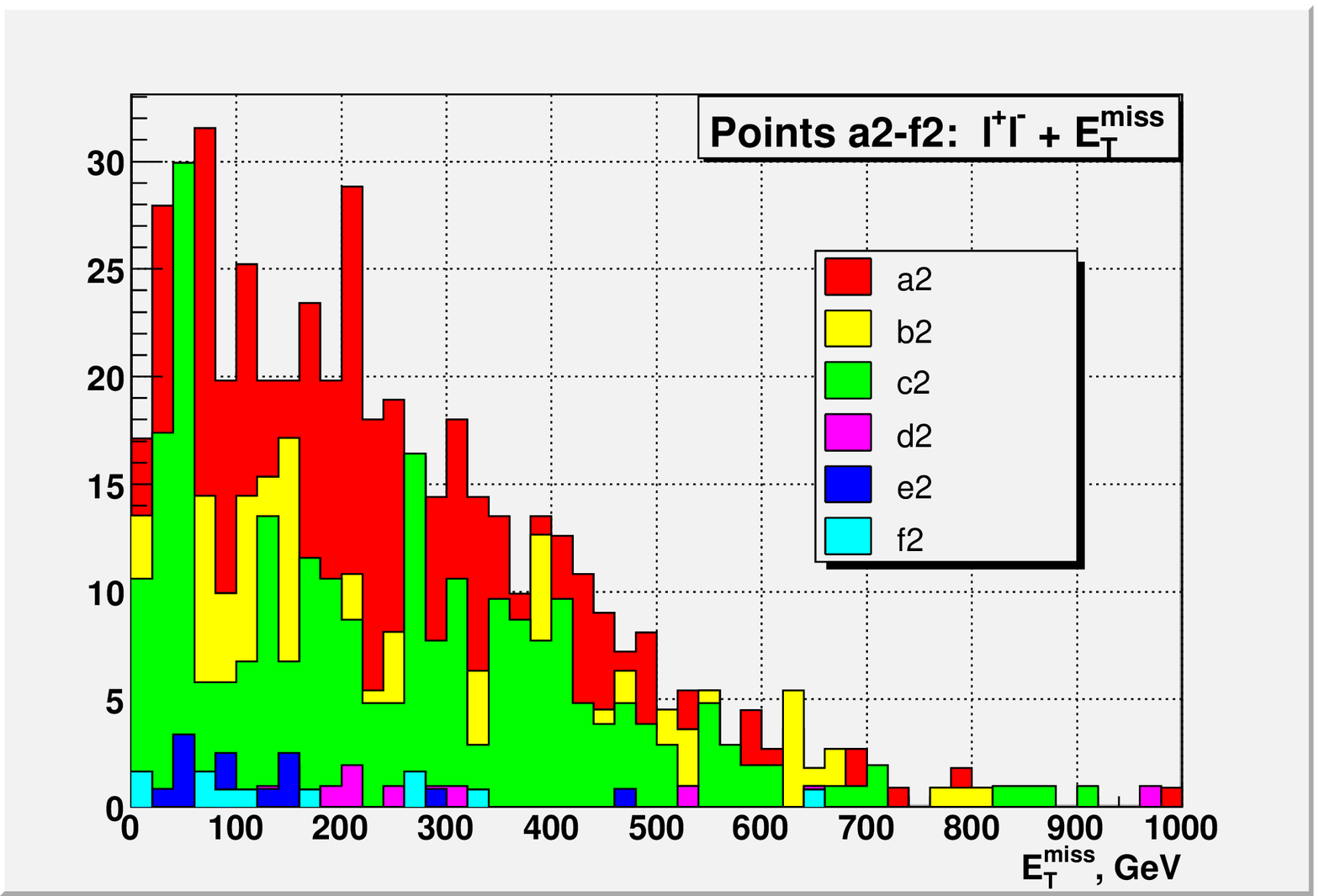}}
    \caption{
             The distributions of $E^{miss}_T$ for points a2 - f2,
             dilepton signature  $ l^+l^- ~+~E^{miss}_T $.
            }
    \label{a2f2dilepton} 
  \end{center}
\end{figure}

\begin{figure}[!htbp]
  \begin{center}
    \resizebox{11.5cm}{!}{\includegraphics{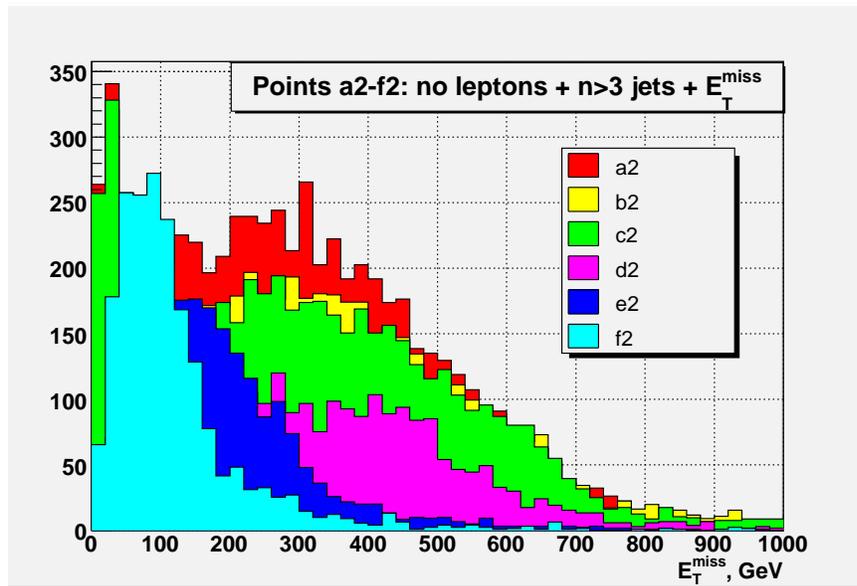}}
    \caption{
             The distributions of $E^{miss}_T$ for points a2 - f2,
             signature $no~leptons ~+~ n \geq ~3~jets ~+~E^{miss}_T $.
            }
    \label{a2f2jets} 
  \end{center}
\end{figure}

\end{document}